# Evaluation of Quantum Annealing-based algorithms for flexible job shop scheduling

Philipp Schworm[a,*], Xiangqian Wu[a], Matthias Klar[a], Jan C. Aurich[a]

[a]*Institute for Manufacturing Technology and Production Systems (FBK), RPTU Kaiserslautern, P.O. Box 3049, 67653 Kaiserslautern, Germany*

* Corresponding author. Tel.: +49-631-205-4066; Fax: +49-631-205-3304. E-mail address: philipp.schworm@rptu.de

**Abstract**

A flexible job shop scheduling problem (FJSSP) poses a complex optimization task in modeling real-world process scheduling tasks with conflicting objectives. To tackle FJSSPs, approximation methods are employed to ensure solutions are within acceptable timeframes. Quantum Annealing, a metaheuristic leveraging quantum mechanical effects, demonstrates superior solution quality in a shorter time compared to classical algorithms. However, due to hardware limitations of quantum annealers, hybrid algorithms become essential for solving larger FJSSPs. This paper investigates the threshold problem sizes up to which quantum annealers are sufficient and when hybrid algorithms are required, highlighting the distribution of computing power in hybrid methods.

## 1. Motivation

The tasks of production planning and control (PPC) encompass scheduling, capacity and quantity-related planning, and monitoring manufacturing and assembly processes within a manufacturing system. PPC aims to ensure on-time production and delivery, consistent capacity utilization, minimized lead times, reduced inventory levels, and enhanced flexibility [1]. However, the landscape of global crises led to rapid and significant changes in the market dynamics of manufacturing enterprises. These changes can result in customer demand fluctuations or short-term material unavailability. As a result, there is a need to develop strategies that ensure the economic sustainability of manufacturing companies. An effective PPC thus emerges as a requirement for the viability of a manufacturing system. Process scheduling, which deals with the sequencing and allocation of jobs, is particularly important within PPC [2]. The objective of process scheduling is to allocate jobs to available resources within the manufacturing system in alignment with its objectives, maximizing efficiency [3]. However, due to the large number of factors to be considered, e.g. multi-objectives and constraints such as dynamic rescheduling, this is a complex optimization problem known as the job shop scheduling problem (JSSP). Although conventional methods like exact algorithms are effective, they often involve long calculation times, which impairs the flexibility of the system and its ability to react to unforeseen events [4]. Consequently, there is a need for computational methods capable of obtaining efficient results in a short time frame. Recent advancements utilizing quantum annealing (QA) have shown promise in addressing this research gap. This research has indicated QA's ability to yield quality solutions for static, dynamic as well as multi-objective JSSPs [5–7]. However, due to the complexity of the problem, hybrid solver configurations and iterative approaches have been utilized. Hybrid methods, while effective, also inherit some of the drawbacks of conventional approaches, such as slower computing times. Iterative methods, while reducing the solution space, may overlook potentially better solutions. To fully harness the potential of QA, it is essential to understand the limitations regarding problem sizes where QA is viable, when hybrid methods become necessary, and when an iterative approach must be employed to achieve solutions within a reasonable timeframe. To address these concerns, various problem instances are defined and computed. Based on the outcomes, an endeavor is made to provide insights into the limitations of QA concerning its application to JSSPs.

The paper is structured as follows: Section 2 outlines the state-of-the-art of solving JSSP and provides an introduction to QA. Furthermore, current QA-based approaches for JSSP will be discussed. After identifying the research gap, Section 3 delineates the structure of the QA-based algorithms (Subsection 3.1). Subsequently, use cases are presented in Subsection 3.2, serving as the basis for the investigations of the QA-based approaches. The experimental results are then detailed in Subsection 3.3. concludes with an outlook and summary.

## 2. State-of-the-art

### 2.1. Job shop scheduling

Job shop scheduling (JSS) is an optimization problem within operations research, frequently applied in manufacturing process scheduling. The formulation of JSS problems varies depending on the constraints and objectives. Classical JSSP aim at allocating given jobs, comprising sequential operations, to manufacturing system machines, considering processing times to achieve specific objectives [3]. These objectives encompass diverse criteria such as time-based, job-number-based, cost-based, revenue-based, and energy- and environment-based factors. Notably, the most prevalent objective is the time-related makespan, minimizing job lead times. In addition, constraints play a pivotal role in mapping manufacturing system conditions to problem formulations [8]. Common constraints include procedure constraints, overlapping constraints, and processing constraints [9]. Assumptions are also made, including uninterrupted machine availability, absence of job priorities and due dates, neglect of transport and setup times, and predetermined operation-machine assignments [8]. However, to better represent real-world scenarios, various JSSPs deviate from these assumptions. Flexible job shop scheduling (FJSSP) assumes machines can perform identical tasks, facilitating variable operation assignments and machine-dependent processing times [9]. Dynamic job shop scheduling (DJSSP) accounts for time-based events like machine failures and probabilistic job arrivals [10]. The complexity of JSSPs, coupled with their NP-hard nature, poses significant challenges for scheduling algorithms [11]. Consequently, obtaining optimal solutions is time-intensive for medium-sized problems and often infeasible for larger instances. Brucker et al.'s findings underline this challenge, demonstrating the computational demands through branch and bound algorithms applied to various JSSP instances [12]. Hence, approximation methods are employed to efficiently compute solutions for larger optimization problems. Approximation techniques include constructive methods, artificial intelligence approaches, local search algorithms, and meta-heuristics [4]. These optimization algorithms are employed individually or in combination to leverage their distinct advantages in addressing diverse JSSP instances. The selection and adaptation of these methods are contingent upon the scale of the problem and the employed modeling techniques. Many approaches for tackling JSSP face a common trade-off: balancing the complexity of the problem or objectives with the need for fast computation, often limiting their applicability to small-scale problems. Exact methods become impractical due to prolonged computation times. While approximate methods can handle larger problem sizes, they either require more computing time or impair the solution quality due to premature termination. Therefore, methods are investigated to navigate this trade-off between computation time and solution quality [13,14]. Recent research in the field of QA-based algorithms also addresses this trade-off by exploring innovative approaches to enhance efficiency without compromising solution accuracy or robustness [5].

### 2.2. Quantum Annealing

The initial realization of QA occurred in 2011 with the development of a cloud-based quantum annealer, also known as an adiabatic quantum computer [15]. Since then, the qubit count, an indicator for the hardware capacity, has experienced exponential growth, progressing from the D-Wave One in 2011 with 128 qubits to the latest D-Wave Advantage boasting 5640 qubits [16]. QA utilizes principles from quantum mechanics to efficiently identify optimal or near-optimal solutions within a minimal timeframe by seeking energy-minimal states of an optimization problem [17]. One method to transform an optimization problem into an energy minimization problem suitable for QA is through the use of the Hamilton function notation. The Hamilton formulation assigns an energy value to each state of the optimization problem, utilizing binary variables $x_i, x_j$ and corresponding scalar weights $Q_{ii}, Q_{ij}$ (Eq. 1).

$$H = \sum_i Q_{ii} x_i + \sum_{i<j} Q_{ij} x_i x_j \qquad (1)$$

To solve the Hamilton function with QA, it's necessary to map it onto the quantum processing unit, a fundamental step for any quantum algorithm utilizing qubits. However, this transformation, known as embedding, can present challenges for quantum annealers, often requiring additional qubits and longer embedding times. Consequently, various embedding techniques, such as auto embedding or external embedding, are employed [18]. As a result, the necessity of overcoming hardware limitations drives the adoption of hybrid solvers and iterative techniques to effectively manage larger problem sizes when solving the Hamilton function with QA. This trend is evident in the literature concerning JSSP computations with QA, where iterative and hybrid solution approaches are already prevalent [7].

### 2.3. Job shop scheduling using Quantum Annealing

In 2015, Venturelli et al. carried out the initial implementation of a JSSP on a quantum annealer [19]. For implementation, the D-Wave two with 509 qubits and the Chimera topology is used. This study delves into the computational efficiency of the D-Wave machine concerning the problem's complexity. To overcome hardware constraints, the researchers employed variable pruning techniques, thereby decreasing the problem sizes which allows computing larger problem instances. Nevertheless, results show that the probability of optimal solutions decreases with increasing problem size. In addition, no upper bound of a specific maximum problem size that can be handled by the quantum annealer is explicitly mentioned. Even in this study, the superiority of QA over classical algorithms in terms of solution quality and efficiency couldn't be conclusively demonstrated, as only small problem sizes were considered. Nonetheless, this work serves as a milestone for numerous subsequent research endeavors in the field. Further investigations have explored limitations associated with constraints on benchmark problems like ft06, employing window shaving techniques to facilitate mapping onto the D-Wave 2000 with the Chimera topology, thus iteratively reducing variables. However, explicit attention to variable bounds was not addressed [20]. Additional research has investigated the boundaries of the JSSP for quadratic instances, reaching up to 15 jobs, operations, and machines on Chimera topologies, and up to 26 jobs, operations, and machines on Pegasus topology on D-Wave Advantage [21]. These studies provide initial indications regarding the limitations of classical QA but without considering classical JSSP with varying processing times and flexible machine assignments in FJSSP. The computation of FJSSP with hybrid solvers was examined in [5]. It was observed that hybrid iterative methods exhibited superior performance in terms of computational time compared to non-iterative hybrids, albeit with a trade-off of partially lower solution quality. Nevertheless, the question of selecting the appropriate solver strategy for different application scenarios remains unanswered. Previous investigations have explored dynamic [6] and multi-criteria objectives [7], relying on iterative and hybrid solution approaches due to problem complexity. However, the limitations of the solvers used persist largely unaddressed.

*2.4. Research gap*

It is unclear when a FJSSP necessitates iterative or hybrid solver approaches for computation. Underutilization of the potential of quantum annealers may result in poorer computation times, while premature adoption of iterative solution methods may restrict the solution space and likely lead to inferior solution quality. For this reason, it is imperative to investigate the relevant limitations and provide insights into performance to make informed decisions.

## 3. Investigations on the limitations of using Quantum Annealing for flexible job shop scheduling

*3.1. Framework and algorithm structure*

The proposed approach aims to investigate the boundaries using QA for solving FJSSPs. Initially, the machines $M$ and jobs $J$ of the FJSSP are combined with the given time range $T$ and transformed into binary variables. These variables lay the foundation for establishing mathematical constraints and objectives for Binary Quadratic Model (BQM). The mathematical model of the FJSSP suitable for QA can be found in [5] and [7], hence these will only be briefly discussed. The BQM formulation is based on binary variables $x_{ijkt} \in X$ equal to 1 if an operation $o_{ij}$ from job $i$ starts at time $t$ and is processed on machine $m_k$. Otherwise, the variable is equal to 0. By linking the variables with corresponding weights (Lagrange parameters) and considering factors such as processing times the BQM is built which is used to construct the Hamiltonian function. The values of Lagrange parameters are determined based on the priority of the objectives and constraints, with constraints typically given higher weights than objective functions. Constraints of the FJSSP are a processing constraint ensuring no multiple starting times of one operation, a procedure constraint so that the predefined sequence of operations is adhered to and an overlapping constraint avoiding multiple operations starting on the same machine. These constraints are enforced through penalty functions, imposing higher energy values in the Hamiltonian for violations. Additionally, the makespan objective is addressed by penalizing deviations beyond a minimum predecessor time. This minimum time is computed by summing the minimum processing times of preceding operations. Finally, the Hamiltonian function is implemented onto quantum hardware and solved using QA-based algorithms. The solver configurations used in this paper are:

- Conventional D-Wave quantum processing unit (QPU) samplers (CQPU)
- Hybrid algorithm using QA, simulated annealing, and tabu search in parallel (HQPU)
- Iterative algorithm using HQPU (IHQPU)

Given that the latest topologies for QA, namely Pegasus and Zephyr, vary in terms of available qubits and connectivity, potentially influencing outcomes, both topologies are applied to each configuration. The iterative solver configuration used is adapted from [7] and shown in Figure 1.

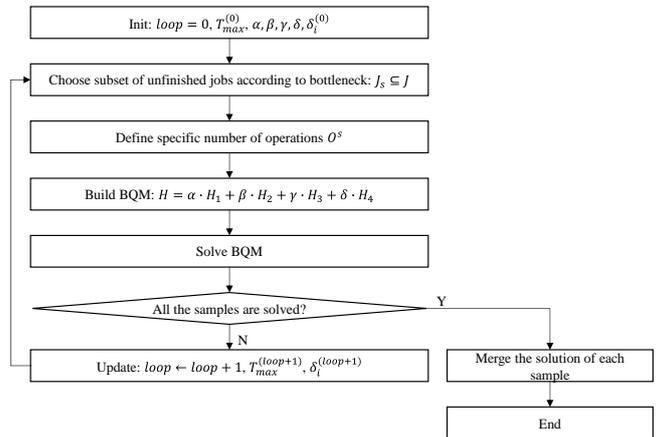

Figure 1: Iterative workflow

In the iterative approach first the Lagrange parameters of the constraints and objective $\alpha, \beta, \gamma, \delta$ and a time range $T$ are determined as well as a bottleneck factor $\delta_0^{(loop)}$ for each job. The bottleneck factors are used to prioritize jobs. In case of makespan jobs are prioritized with high total processing times that only have a small selection of machines. An initial set of jobs $J$ is divided into a smaller subset of jobs $J_s$ according to the bottleneck factors. For each loop a suitable number of operations $O_s$ is chosen. Afterwards the BQM of the corresponding subproblems is built and solved with the chosen solver. This is repeated for each subproblem, with the bottleneck factors and time ranges are adjusted accordingly. When all subproblems are solved then the solutions of all loops are merged.

## 3.2. Use cases

The solver configurations are applied to FJSSPs to evaluate their limitations. These investigations are structured into three experimental setups, each aimed at exploring different aspects of these limitations. In the *first experimental setup*, inspired by Carugno et al.'s work, quadratic instances are utilized [21]. In these instances, JSSPs are characterized by scaling the number of operations per job $O$ with the number of machines and number of jobs. In addition, the given time range $T_r$ including the discrete times when each operation can be processed is varied. Each operation's processing $p_{ij}$ time is standardized to 1. Consequently, optimal solutions can be determined, allowing for an examination of the maximum number of processable variables in the BQM in combination with consideration of the computing times for each solver. The *second experimental setup* involves FJSSP by varying the number of feasible machines $k$ for each operation while maintaining constant processing times and constant $T_r$. This design enables an investigation into how the number of quadratic interactions in the BQM affects solution quality and processable problem sizes. In the *third experimental setup*, processing times $p_{ij}$ and $T_r$ are varied. This variation facilitates an exploration of how changes in processing times influence the number of processable jobs and operations. Detailed information on the experimental setups can be found in Table 1. The computations are performed on a XEON_SP_6126 with 20 GB RAM and the respective quantum hardware. A time limit of 15 minutes is set for achieving a solution with the respective solver configuration.

Table 1: Experimental setup

| Experimental Setup | J=O=M | k | p | T | $T_r$ |
|---|---|---|---|---|---|
| 1 | 1-100 | 1 | 1 | 2-102 | 2-5 |
| 2 | 2-25 | 2-25 | 1 | 3-26 | 2 |
| 3 | 2-15 | 2-15 | 2-15 | 3-226 | p+1 |

## 3.3. Comparison and results

Selected computational outcomes focusing on its limitations are summarized in Table 2. Various parameters are documented for evaluation purposes. These include the specific solver configurations, computation times $T_c$, which denote the duration from problem submission to solution retrieval. Additionally, solution quality can be assessed through the achieved makespan $ms$. Furthermore, each problem's characteristics are described, encompassing the number of variables $n_v$ and quadratic interactions $n_q$ within the BQM, along with the number of qubits employed for embedding $n_e$. In each line six values are given for $T_c$, and $ms$ which represent the results of CQPU-Zephyr, CQPU-Pegasus, HQPU-Zephyr, HQPU-Pegasus, IHQPU-Zephyr, and IHQPU-Pegasus, where a dash implies that no solution could be found within the given time interval or due to hardware limitations. $n_e$ is recorded only for CQPU-Zephyr and CQPU-Pegasus. It is important to emphasize that the solutions generated by IHQPU partially match those of HQPU, depending on the problem size. This is since the parent problem is only partitioned if it exceeds a certain size threshold. For evaluation, the results of experimental setup 1 with $T_r = 2$, experimental setup 2 with J = O = M = k, and experimental setup 3 with $T_r = 2$ and J = O = M = k = p are visualized in Figure 2-4. Further results can be found in the supplementary materials.

The computational results show the advantages of the Zephyr topology over the Pegasus topology. In experimental setup 1 with $T_r = 2$, both CQPU-Zephyr and CQPU-Pegasus achieve optimal makespans, with CQPU-Zephyr exhibiting shorter computation times. This discrepancy likely arises from the greater availability of connected qubits in Zephyr topologies, resulting in less significant increases in problem size during embedding compared to Pegasus. However, Zephyr's potential is limited by its fewer available qubits, restricting its solvable problem size to 20 for CQPU-Zephyr, whereas CQPU-Pegasus can handle problems up to size 32. The HQPU results demonstrate advantages in manageable problem sizes, attributable to additional CPU resources for computation. Here HQPU-Zephyr can handle problem sizes up to 86 whereas HQPU-Pegasus achieves feasible solutions up to 84 in the 15-minute time frame. This is because the

Table 2: Computational results

| J = O = M | k | p | T | $T_r$ | $n_v$ | $n_q$ | $T_c$ in s[1] | $ms$[1] | $n_e$[2] |
|---|---|---|---|---|---|---|---|---|---|
| 20 | 1 | 1 | 21 | 2 | 800 | 1160 | 14.4/29.9/6.8/12.5/6.8/12.5 | 20/20/20/20/20/20 | 1021/1372 |
| 32 | 1 | 1 | 33 | 2 | 2048 | 3008 | -/207.4/14.9/18.2/14.9/18.2 | -/32/32/32/32/32 | -/3191 |
| 84 | 1 | 1 | 85 | 2 | 14112 | 21000 | -/-/829.4/676.3/225.8/158.2 | -/-/84/84/84/84 | -/- |
| 86 | 1 | 1 | 87 | 2 | 14792 | 22016 | -/-/864.8/-/144.2/241.1 | -/-/86/-/86/86 | -/- |
| 45 | 1 | 1 | 46 | 2 | 4050 | 5985 | -/-/44.3/51.7/42.1/57.6 | -/-/45/45/45/45 | -/- |
| 49 | 1 | 1 | 50 | 2 | 4802 | 7105 | -/-/64.5/69.3/49.8/52.1 | -/-/50/50/50/50 | -/- |
| 19 | 1 | 1 | 20 | 2 | 648 | 936 | 9.53/13.3/6.7/10.4// | 19/19/19/19/19/19 | 998/1192 |

[1] CQPU-Zephyr/CQPU-Pegasus/HQPU-Zephyr/HQPU-Pegasus/IHQPU-Zephyr/IHQPU-Pegasus

[2] CQPU-Zephyr/CQPU-Pegasus

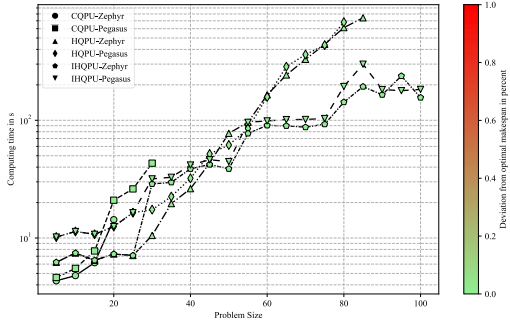
Figure 2: Experimental setup 1 with J=O=M

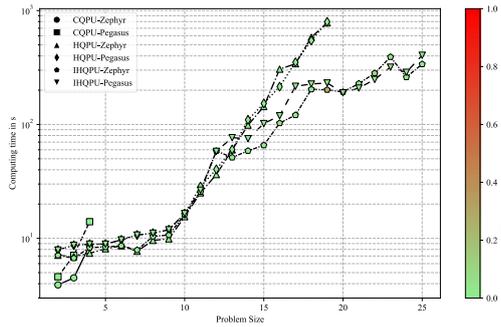
Figure 3: Experimental setup 2 with J=O=M=k

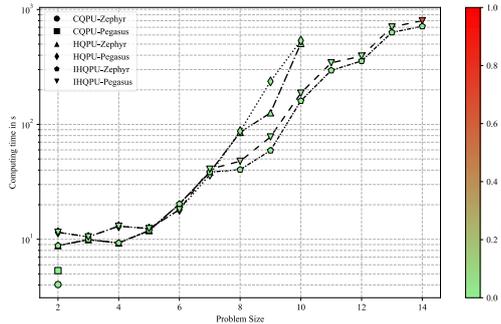
Figure 4: Experimental setup 3 with J=O=M=k=p

HQPU-Zephyr configuration consistently exhibits shorter computing times than HQPU-Pegasus due to the lower embedding effort.

A direct comparison between HQPU and CQPU shows that CQPU computations initially outperform HQPU until reaching problem sizes of 19 for Pegasus and for Zephyr. Upon reaching the range of the previously described thresholds, HQPU outperform CQPU. This behavior can also be explained by the embedding process, which is less time-consuming in hybrid methods since only a subproblem is embedded and computed on the QPU, while the rest is processed on the CPU due to the previous decomposition. Applying IHQPU configurations shortens the computing times even further, especially for large problems. Thus, the IHQPU-Pegasus method achieves better computing times beginning from 49 and the IHQPU-Zephyr from 45 with consistent solution quality. It's worth highlighting that the IHQPU-Zephyr configuration, for instance, achieves an optimal solution for a problem size of 100 in just 150 seconds, whereas the HQPU-Zephyr configuration only manages a problem size of 87 within 15 minutes. As problem sizes approach the requiring approximately 30 seconds with the hybrid method, it appears prudent to adopt an iterative approach. Beyond this threshold, the exponential increase in computing time becomes more evident. Similar results apply to changing $T_r$ and to other types of problems and experimental setups.

In experimental setup 2, the increase in quadratic interactions results in a significant reduction in computable problem sizes. While focusing on problem instances with J = O = M = k CQPU-Zephyr and CQPU-Pegasus fail to find embeddings starting from 5. Even for smaller problem sizes, the computation times of the HQPU configurations are superior. However, a deterioration in solution quality can be observed by computation of some problem instances from 16 for IHQPU-Zephyr and 20 for IHQPU-Pegasus. Beyond this size, it seems reasonable to consider the trade-off between solution time and computational quality. Additionally, CQPU-Pegasus increasingly yields non-optimal solution qualities. However, this phenomenon is exclusive to CQPU configurations and not observed with HQPU. The longer chain lengths occurring during embedding due to numerous quadratic interactions contribute to this issue, whereas CQPU-Zephyr produces fewer non-optimal solutions due to its higher qubit connectivity.

Experimental setup 3 yields similar conclusions, as the varying processing time results in increased variables and quadratic interactions, consequently diminishing the computable problem size. In experimental setup 3, however, non-optimal solutions occur more frequently during the iterative process. This increased frequency is probably due to the expanded solution space, which leads to smaller differences in the weighting of the individual solutions. Generally, only a minority of computations fail to identify an optimal makespan. However, considering the frequent occurrence of non-optimal solutions in prior studies, it suggests a correlation between effective solutions and problem structures, [7]. Consequently, the next phase involves examining the behavior of individual solver configurations within industrial settings, emphasizing the significance of hybrid and iterative approaches in addressing industrial-scale problems.

## 4. Summary and outlook

In this paper, the performance of various QA-based solver configurations across multiple instances of JSSP and FJSSP problems is investigated. These configurations leverage QPU topologies, namely Pegasus and Zephyr, in CQPU, HQPU, and IHQPU. It was found that configurations utilizing the Zephyr topology generally exhibit shorter computation times compared to those employing Pegasus, albeit constrained by hardware limitations in handling larger problem sizes. The performance of the embedding process is notably affected, particularly as the problem size nears the hardware's capacity limits. Hybrid approaches demonstrate improved computation times in these situations. However, as problem sizes increase, rising processing times are also encountered by the HQPU. Consequently, employing IHQPUs becomes beneficial beyond a certain problem size threshold, albeit potentially compromising solution quality. However, since the problems under consideration are of a theoretical nature, it makes sense to extend the investigations. In order to enable a holistic evaluation, future research aims to extend these investigations into industrial settings, encompassing diverse multi-criteria objectives and substantial disparities in processing times.

## Acknowledgements

This research was funded by the Ministerium für Wirtschaft, Verkehr, Landwirtschaft und Weinbau Rheinland-Pfalz - 4161-0023#2021/0002-0801 8401.0012.